\documentclass[5p,times]{elsarticle}
\usepackage[english]{babel}

\usepackage[dvipsnames]{xcolor}
\usepackage{tikz}
\usepackage{amsmath}
\usepackage{subcaption}
\usepackage[hyphens]{url}

\usepackage{xstring}
\usepackage{xspace}
\newcommand\reduce{\textit{reduce}\xspace}
\newcommand\broadcast{\textit{broadcast}\xspace}
\newcommand\canary{\textsc{Canary}\xspace}
\usepackage{makecell}
\usepackage{soul} %
\usepackage{hhline}
\usepackage{enumitem}
\usepackage{mdframed}

\usepackage{graphicx}
\usepackage{capt-of}%
\usepackage{booktabs}
\usepackage{varwidth}
\usepackage{fontawesome}
\newcommand{\reva}[1]{{#1}}
\newcommand{\revb}[1]{{#1}}

\newsavebox\tmpbox

\definecolor{mygreen}{HTML}{63ACBE}
\definecolor{myred}{HTML}{EE442F}
\definecolor{myyellow}{HTML}{601A4A}

\newcommand{\up}{\textcolor{mygreen}{\faCheck}}
\newcommand{\down}{\textcolor{myred}{\faClose}}
\newcommand{\half}{\textcolor{myyellow}{\faSearch}}

\newcommand{\uptext}{\textcolor{mygreen}{\faCheck}}
\newcommand{\downtext}{\textcolor{myred}{\faClose}}
\newcommand{\halftext}{\textcolor{myyellow}{\faSearch}}

\journal{Future Generation Computer Systems}

\begin{document}

\begin{frontmatter}

\title{Canary: Congestion-Aware In-Network Allreduce Using Dynamic Trees
}

\author[eth,sapienza]{Daniele De Sensi\corref{cor1}}
\ead{desensi@di.uniroma1.it}

\author[ethee]{Edgar Costa Molero}
\ead{cedgar@ethz.ch}

\author[eth]{Salvatore Di Girolamo}
\ead{salvatore.digirolamo@inf.ethz.ch}

\author[ethee]{Laurent Vanbever}
\ead{lvanbever@ethz.ch}

\author[eth]{Torsten Hoefler}
\ead{torsten.hoefler@inf.ethz.ch}

\cortext[cor1]{Corresponding author}

\affiliation[eth]{organization={Dept. of Computer Science, ETH Zurich},%
            addressline={Rämistrasse 101}, 
            city={Zürich},
            postcode={8092}, 
            country={Switzerland}
}
\affiliation[ethee]{organization={Dept. of Information Technology and Electrical Engineering, ETH Zurich},%
            addressline={Rämistrasse 101}, 
            city={Zürich},
            postcode={8092}, 
            country={Switzerland}
}
\affiliation[sapienza]{organization={Dept. of Computer Science, Sapienza University of Rome},%
            addressline={ Piazzale Aldo Moro 5}, 
            city={Rome},
            postcode={00185}, 
            country={Italy}
}

\begin{abstract}
The allreduce operation is an essential building block for many distributed applications, ranging from the training of deep learning models to scientific computing. In an allreduce operation, data from multiple hosts is aggregated together and then broadcasted to each host participating in the operation. Allreduce performance can be improved by a factor of two by aggregating the data directly in the network. Switches aggregate data coming from multiple ports before forwarding the partially aggregated result to the next hop. In all existing solutions, each switch needs to know the ports from which it will receive the data to aggregate. However, this forces packets to traverse a predefined set of switches, making these solutions prone to congestion. For this reason, we design \canary, the first congestion-aware in-network allreduce algorithm. \canary uses load balancing algorithms to forward packets on the least congested paths. Because switches do not know from which ports they will receive the data to aggregate, they use timeouts to aggregate the data in a best-effort way. We develop a P4 \canary prototype and evaluate it on a Tofino switch. We then validate \canary through simulations on large networks, showing performance improvements up to 40\% compared to the state-of-the-art.
\end{abstract}


\end{frontmatter}

\section{Introduction}\label{sec:introduction}
As the parallelism in computing systems steadily increases, the performance scalability of applications running on data centers becomes more dependent on communication performance. The allreduce operation is a widely-used communication primitive, both for the training of machine learning models~\cite{ddl_survey}, but also in scientific computing in general~\cite{milc,fftw}. In an allreduce, each host has a vector of data elements. All the vectors must be \textit{reduced} (i.e., aggregated) together element-wise using a commutative and associative operator. Then, after aggregation, data is distributed back to the hosts. 

Allreduce accounts for a significant fraction of the training time of deep learning models, with estimates ranging from 50\% for 10 Gbps networks~\cite{switchml}, to 20-30\% for 100 Gbps networks~\cite{nvidiaallreduce}. Moreover, improvements in computation speed significantly outpace network bandwidth improvements. For example, we observed a 10x increase in GPU floating-point performance in 2.5 years~\cite{nvidiaa100,switchml}. In contrast, it took ten years for network bandwidth to increase by 10x~\cite{ibspeeds,switchml}. Thus, we can expect application performance to be even more dependent on communication performance in the future.

For these reasons, several allreduce optimization techniques have been proposed~\cite{ddl_survey}, including (but not limited to) data quantization~\cite{quantization}, sparsification~\cite{sparsification}, non-blocking collectives~\cite{nbc,async}, solo allreduce~\cite{solo_allreduce}, \reva{hierarchical synchronization~\cite{10.1145/3229543.3229544,9565148}}, and in-network reductions~\cite{switchml,atp,sharp,nvidiaallreduce}. This work focuses on in-network reductions, i.e., solutions where the network switches aggregate data. Several works showed that in-network allreduce transmits half of the data volume transmitted by the host-based bandwidth-optimal allreduce algorithm \cite{bwoptimalallreduce} (e.g., \textit{ring} allreduce)~\cite{atp,switchml,nvidiaallreduce}. Thus, if the network aggregates the data at line rate, this potentially halves the time required to complete the reduction. 

\begin{figure}[htpb]
\begin{subfigure}{.47\columnwidth}
  \centering
    \includegraphics[width=\linewidth]{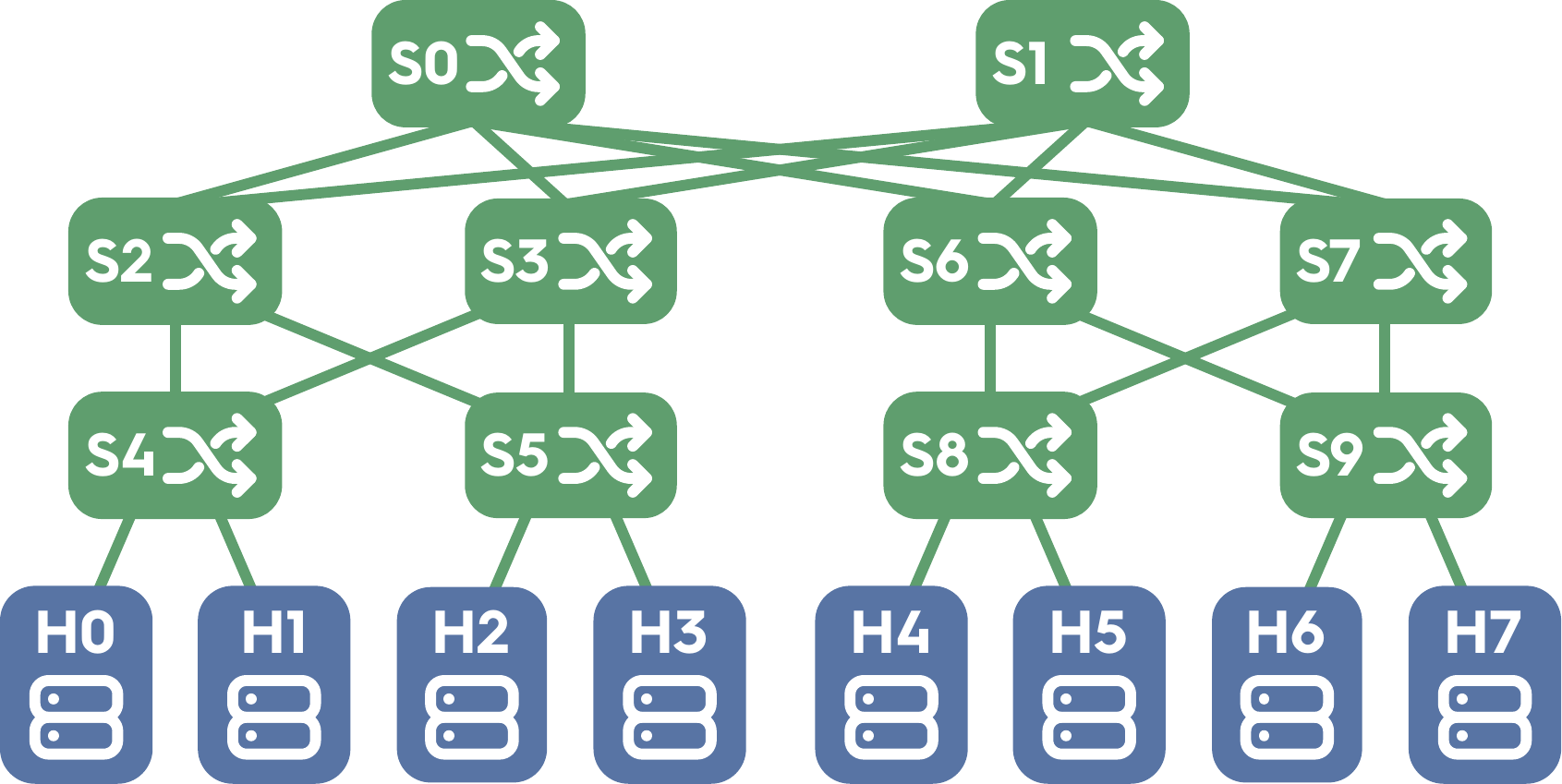}
    \caption{A network with 8 hosts.}
    \label{fig:motivation:new:a}
\end{subfigure}%
\hfill
\begin{subfigure}{.47\columnwidth}
  \centering
    \includegraphics[width=\linewidth]{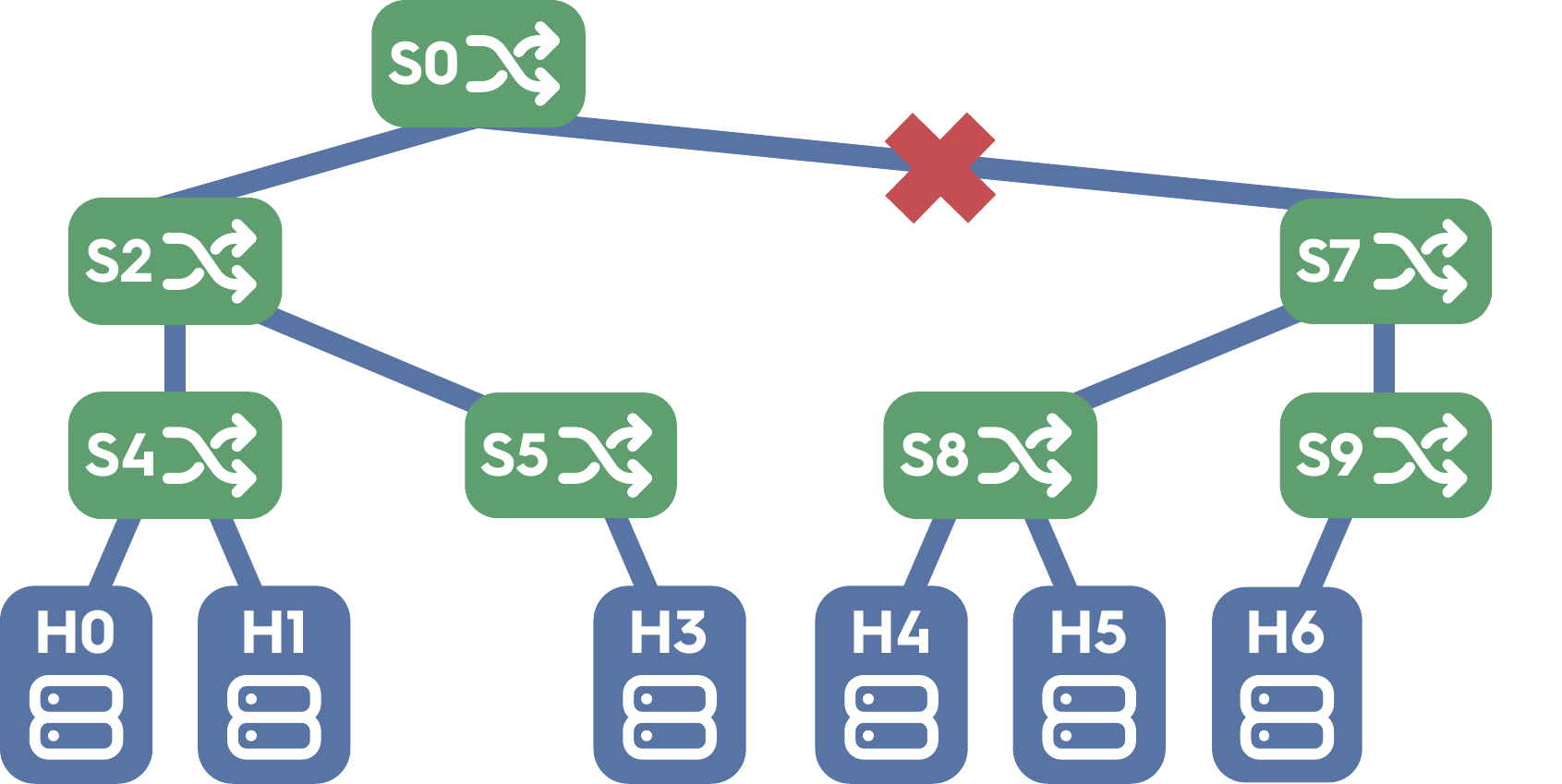}
    \caption{A reduction tree.}
    \label{fig:motivation:new:b}
\end{subfigure}
\caption{In-network allreduce example.}
\label{fig:motivation:new}
\end{figure}

All existing in-network allreduce algorithms adopt a similar approach~\cite{switchml,atp,panama,sharp,sharp2,nvidiaallreduce}, which we describe through an example. Let us consider the network depicted in Figure~\ref{fig:motivation:new:a}, where the hosts \textit{H0}, \textit{H1}, and \textit{H3-H6} want to perform an in-network allreduce. First, they set up a \textit{reduction tree}, where the leaves are the hosts participating in the reduction, and the intermediate nodes are a subset of the switches in the network, as shown in Figure~\ref{fig:motivation:new:b}. This setup step mostly involves installing forwarding rules in the switches. By doing so, for example, \textit{{S4}} knows that it must aggregate the data coming from \textit{{H0}} and \textit{{H1}}, and forward the aggregated result to \textit{{S2}}. Similarly, \textit{S5} does not wait for data coming from \textit{H2}, and forwards the data coming from \textit{H3} to \textit{S2} immediately after receiving it.

Albeit the described algorithm is simple and effective, it is also prone to congestion, because each switch in the reduction tree needs to know a priori its children and its parent in the reduction tree. This forces packets to be always routed on the same paths, regardless of their congestion. Network congestion can significantly slow down applications~\cite{10.1145/3570609,246494,slingshot,netnoise,8049022}, and this is particularly relevant for in-network reductions. For example, let us assume that the link between \textit{S7} and \textit{S0} in Figure~\ref{fig:motivation:new} is congested. Even if \textit{S0} already received the data from all its other children, it still needs to wait for the data coming from \textit{S7} before starting the \broadcast phase. 

Thus, \textbf{it is enough to have congestion on just one of the links composing the reduction tree to slow down the entire operation}. A straightforward solution would consist in running the allreduce traffic in a separate traffic class. However, as we also show in Section~\ref{sec:eval:concurrent}, concurrent allreduces issued by different applications (e.g., different training jobs) would still interfere unless they are mapped to different classes. Because the number of concurrent allreduces can be higher than the available traffic classes~\cite{atp,slingshot,qos-1,qos-2}, this is not a viable solution.

For these reasons, in this work \textbf{we design and evaluate \canary (\textit{\underline{C}ongestion-\underline{A}ware In-\underline{N}etwork \underline{A}ll\underline{r}educe Using D\underline{y}namic Trees}), the first congestion-aware in-network allreduce algorithm}. \canary relies on traffic load balancing algorithms to send packets on the least congested paths, dynamically building the reduction tree and adapting it throughout the execution based on congestion.

\begin{figure}[htpb]
    \centering
    \includegraphics[width=\columnwidth]{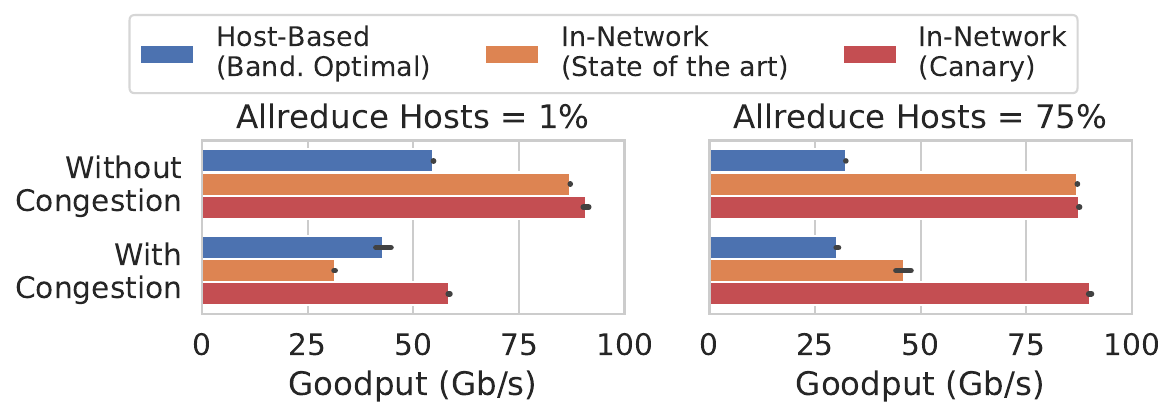}
    \caption{Goodput of the bandwidth-optimal host-based allreduce, of the state-of-the-art in-network allreduces, and of \canary, when running on 1\% and 75\% of the hosts in a 1024 hosts network.}
    \label{fig:motivation:exp}
\end{figure}

To illustrate the impact of congestion on in-network allreduce, we simulate a 2-level fat tree network~\cite{fattree} connecting 1024 hosts (we provide more details on the simulation infrastructure in Section~\ref{sec:evaluation:sst}). We execute an allreduce first on 1\% and then on 75\% of the hosts in the network. We observe in Figure~\ref{fig:motivation:exp} that, when there is no congestion, both state-of-the-art in-network allreduce algorithms (using a static reduction tree) and \canary provide a 2x bandwidth improvement over the bandwidth-optimal host-based allreduce. 

Then, we introduce congestion by concurrently executing a random uniform injection traffic pattern on the remaining hosts (99\% and 25\%, respectively). We observe that congestion causes a drop in the performance of the state-of-the-art in-network allreduce, which can even perform worst than the host-based allreduce. 

Instead, \canary is less affected by congestion and provides a performance advantage over both the bandwidth-optimal host-based allreduce and state-of-the-art allreduces. As we describe in detail in Section~\ref{sec:design}, this is possible because \canary dynamically builds and adapts the reduction tree to the network conditions by relying on existing congestion-aware traffic load balancing techniques.

In this work, we introduce the following contributions:
\begin{itemize}[leftmargin=*]
\item We identify the impact of congestion on in-network allreduce, and we design an algorithm that relies on dynamic in-network reduction trees (Section~\ref{sec:design}).
\item We improve the management of the switch resources and the fault tolerance compared to the state-of-the-art because the switches only store a soft state (Section~\ref{sec:design}).
\item We implement a P4 \canary prototype on a Tofino switch~\cite{tofino}, to assess the feasibility and limitations of our algorithm (Section~\ref{sec:implementation} and Section~\ref{sec:discussion}).
\item We perform a detailed analysis through large-scale simulations, calibrated on our P4 implementation (Section~\ref{sec:evaluation}) showing that, on congested networks, \canary is up to 40\% faster than state-of-the-art in-network allreduce algorithms.
\end{itemize}

\section{State of the Art}\label{sec:motivation}
We now discuss \canary's fundamental design principles (summarized in Table~\ref{tab:state}), that distinguish it from most existing in-network allreduce algorithms.

\begin{table}[htpb]
\footnotesize
\centering

\begin{tabular}{|l|c|c|c|c|} 
 \hline
 {\textsc{\makecell{Algorithm/\\Network}}} & \textsc{Year} & {\textsc{\makecell{CA}}} & {\textsc{\makecell{DRM}}} & {\textsc{\makecell{DFT}}}  \\
 \hline\hline
 PERCS~\cite{percs}           & 2010 & \down & \faQuestion & \faQuestion \\ \hline
 Aries~\cite{aries}           & 2012 & \down & \faQuestion & \faQuestion \\ \hline
 Tofu~\cite{tofu}             & 2018 & \down & \faQuestion & \faQuestion \\ \hline
 SHARP~\cite{sharp,sharp2}    & 2020 & \down & \down & \down \\ \hline
 Klenk et al.\cite{nvidiaallreduce} & 2020 & \down & \up   & \down \\ \hline
 PANAMA~\cite{panama}         & 2020 & \half & \up & \down \\ \hline
 ATP~\cite{atp}               & 2021 & \down & \up   & \down \\ \hline
 SwitchML~\cite{switchml}     & 2021 & \down & \down & \down \\ \hline
 OmniReduce~\cite{omnireduce} & 2021 & \down & \down & \down \\ \hline
 PIUMA~\cite{hyperx-innet}    & 2021 & \down & \faQuestion & \down \\ \hline
 Flare~\cite{flare}           & 2021 & \down & \down & \down \\ \hline
 \textbf{\textsc{Canary}}     & 2022 & \up   & \up   & \up   \\ \hline
\end{tabular}

\caption{\canary design principles, and comparison with the state-of-the-art. \up: considered, \half: partially considered, \down: not considered, \faQuestion: unknown. \textbf{\textsc{CA}}: Congestion-Awareness, \textbf{\textsc{DRM}}: Dynamic Resource Management, \textbf{\textsc{DFT}}: Dynamic Fault Tolerance.}
\label{tab:state}
\end{table}

\subsection{Congestion Awareness}\label{sec:motivation:ca}
Existing interconnection networks have a large path diversity~\cite{fatpaths,slingshot} and, to avoid congestion, load balance the traffic using algorithms like ECMP~\cite{ecmp}, flowlet switching~\cite{conga,201562}, Valiant routing~\cite{valiant}, and others. Some of these algorithms, like ECMP, distribute packets over the available paths by selecting the destination port based on the result of a hash function computed on some packet header fields. However, although many networks use ECMP~\cite{10.1145/2829988.2787508}, it has been shown that traffic often experiences congestion, even in the presence of alternative non-congested paths~\cite{conga,drill,10.1145/2535372.2535375}. For this reason, some load balancing algorithms try to select the least congested path among those available. Such algorithms are also offered by some of the largest cloud providers when deploying high-performance virtualized clusters~\cite{srd,azurear}. 

As discussed, however, all state-of-the-art in-network reduction algorithms always send the packets on the same paths regardless of congestion (Table~\ref{tab:state}, \downtext). Moreover, to the best of our knowledge, only one in-network allreduce algorithm distributes the traffic over multiple reduction trees~\cite{panama}, showing advantages compared to a single reduction tree (Table~\ref{tab:state}, \halftext). Nevertheless, because the algorithm statically selects trees in a round-robin way, it is still congestion oblivious, even if it balances traffic over multiple paths.

Differently from all existing solutions, \canary dynamically builds, packet by packet, the optimal reduction tree based on the current network status, by routing packets on the least congested paths, and by aggregating the packets in a best-effort fashion (Table~\ref{tab:state}, \uptext -- Section~\ref{sec:design:general}). 
It is worth remarking that simply adding load balancing capabilities to existing in-network allreduce algorithms would not work and that we need to design new ways to aggregate the data, as we discuss in Section~\ref{sec:design}.

\subsection{Dynamic Resource Management}

Each switch aggregates in a memory buffer, packet by packet, the data it receives from its children. Switches, however, have limited memory, and existing in-network allreduce algorithms adopt different approaches to deal with this. Most algorithms reserve some buffer space before starting an allreduce and, if there is no buffer space available, they fall back to a host-based algorithm~\cite{sharp,sharp2,tofu,daiet}. Because this reservation process requires interacting with the control plane and might introduce some latency (up to \textit{``a few seconds''}~\cite{panama}), resources are usually reserved when the application starts and deallocated when the application terminates~\cite{sharp,sharp2,switchml}. 

However, by doing so, long-lived applications would reserve resources for their entire execution, even if they would only sporadically use in-network reductions, potentially excluding other applications from using them. This decreases the number of concurrent in-network reductions that can be executed, which might be a relevant problem on multi-tenant datacenters~\cite{atp} (Table~\ref{tab:state}, \downtext). Other approaches, instead, dynamically partition the available resources across the currently active reductions~\cite{panama,nvidiaallreduce,atp}, as we also do in \canary (Table~\ref{tab:state}, \uptext -- Section~\ref{sec:design:collisions}). Specifically, in \canary the switches allocate and deallocate the memory in an on-demand fashion and strictly for the time required to complete the reduction.

\subsection{Dynamic Fault Tolerance}\label{sec:limitations:fault}
Another critical point to discuss is how to deal with links or switch failures~\cite{10.1145/3278532.3278566,10.1145/3464994.3464996}. Because existing algorithms statically determine the reduction trees, if a link or a switch on the tree fails, the in-network reduction cannot progress. In most cases, the network controller detects switch failures~\cite{sharp,panama} and builds another reduction tree. However, hosts might be in an inconsistent state after a failure (e.g., some hosts might have successfully received all the reduced data, whereas others might just have received part of the data). Recovering from such a state might imply re-issuing the entire reduction operation~\cite{sharp,sharp2} from scratch on a different reduction tree or falling back to host-based reductions (Table~\ref{tab:state}, \downtext). Other solutions~\cite{switchml} delegate the task of detecting and recovering from failure to the upper layer. 

\canary, on the other hand is self-contained and can autonomously detect and recover from switch failures without re-starting the entire reduction from scratch. Indeed, \canary builds reduction trees dynamically and keeps only a soft state in the switches, and treats switches and links failures in the same way as a packet loss. In both cases, \canary requires only the retransmission of the small fraction of data that was stored in the switch when it failed (Table~\ref{tab:state}, \uptext -- Section~\ref{sec:design:losses}). Because managing both packet losses and switches faults adds complexity to the algorithm, \canary partitions its functionalities between switches and a \textit{leader host} (Section~\ref{sec:design:leader}).

\section{\textsc{Canary} Design}\label{sec:design}
In general, we can consider in-network allreduce algorithms composed of two phases: a \reduce phase (where data flows from the hosts to the root of the reduction tree) and a \broadcast phase (where aggregated data flows from the root to the hosts). In \canary, hosts send packets to the same root switch (predetermined before starting the application), but packets are forwarded on the least congested paths towards the root to bypass congestion. \canary is orthogonal to the load balancing algorithm, and switches can use any existing algorithm to select the next-hop (either on a per-packet~\cite{drill} or a per-flowlet granularity~\cite{conga}). 

\canary aggregates packets that traverse the same switch in the same time window. Each switch allocates memory on-demand when receiving packets in the \reduce phase and deallocates it in the \broadcast phase, after sending the aggregated data to its children. To simplify the description of the algorithm, we first describe in Section~\ref{sec:design:general} a scenario where switches have infinite memory, where a fully reliable network is used (i.e., packets are never dropped and switches/links never fail), and where there is at most one application at a time using \canary. We then remove these assumptions in Section~\ref{sec:design:collisions}, Section~\ref{sec:design:losses}, and Section~\ref{sec:design:multitenancy} respectively.

\subsection{General Design}\label{sec:design:general}
Before describing the details of the algorithm, we analyze the challenges of using non-predetermined paths in in-network reductions. For the moment, we assume that each host can fit all the data it needs to reduce in a single network packet. We then discuss in Section~\ref{sec:design:general:packets} how to deal with larger data.  

\begin{figure}[t]
    \centering
    \includegraphics[width=\columnwidth]{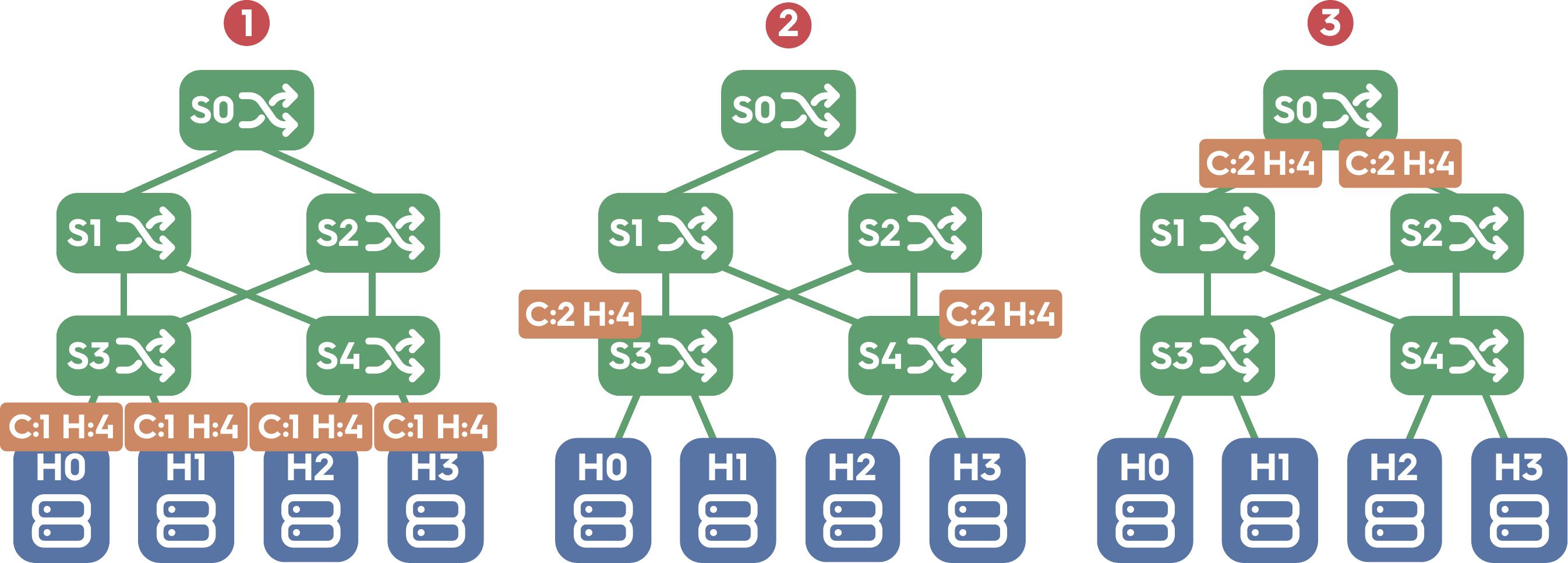}
    \caption{Aggregation counter update. \textit{C:} aggregation counter, \textit{H:} number of hosts.}
    \label{fig:design_topology}
\end{figure}

If we consider the network shown in Figure~\ref{fig:design_topology}, we can see that if we select \textit{{S0}} as a root and we let packets follow different paths from the hosts to the root, \textit{{S1}} and \textit{{S2}} would not know how many packets they will receive from their children. According to the network conditions, sometimes both \textit{{S3}} and \textit{{S4}} might decide to forward their packets to \textit{{S1}}, sometimes they might decide to forward their packets to \textit{{S2}}, and sometimes \textit{{S3}} might forward packets to \textit{{S1}} and \textit{{S4}} to \textit{{S2}} (or vice-versa). As a consequence, \textit{{S1}} might receive 0, 1 or 2 packets to aggregate (and the same for \textit{{S2}}). This ambiguity is not present in existing in-network reduction approaches because packets always follow predetermined paths, and each switch knows exactly how many packets to wait for and aggregate. For this reason, existing in-network allreduce algorithms cannot simply be extended by using congestion-aware traffic load balancing, and a different approach must be adopted.

\subsubsection{Reduce}\label{sec:design:general:reduce}
Because the switch does not know how many packets to wait for, in the \reduce phase \canary aggregates all packets received in a given time window. The first time a switch receives a reduction packet, it creates a \textit{descriptor} and stores it in its memory. The descriptor contains a data accumulator (where the switch stores the data carried by the packet) and the root address (also carried by the packet). The descriptor also contains a timer that the switch starts when the first reduction packet is received. After storing the information carried by the packet in the descriptor, the switch drops the packet. The switch stores in the descriptor also the list of ports from which it received the allreduce packets (it will use it to reach the children in the \broadcast phase). For all subsequent packets, the switch aggregates the data carried in the packets with that stored in the accumulator and then drops the packets. When the timeout expires, the switch retrieves the accumulated data from the descriptor, stores it in a new packet, and sends it to the next hop towards the root. The switch selects the next hop using any available congestion-aware load balancing algorithm, thus dynamically building the reduction tree depending on the current network conditions. 

Eventually, all packets reach the root and the \reduce phase is concluded. Each packet sent by the hosts carries a counter indicating the number of already reduced packets, together with the number of hosts participating to that reduction (Figure~\ref{fig:design_topology} (\includegraphics[scale=0.031,trim=0 40 0 0]{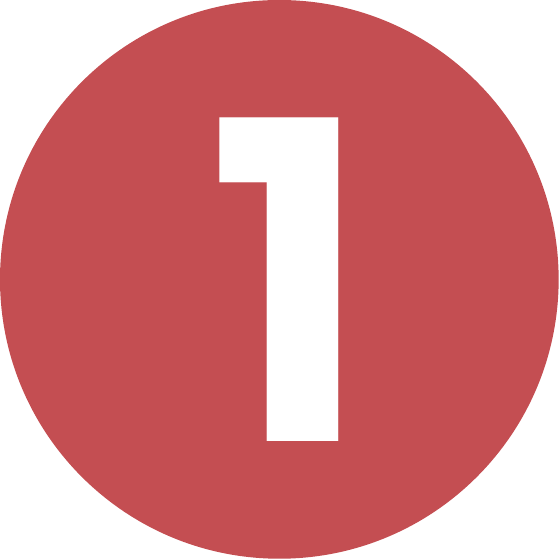})). Counters coming from multiple packets are summed by the switches and stored in the descriptor. For example, if \textit{{S3}} reduces the data coming from \textit{{H0}} and \textit{{H1}}, it will send a packet to \textit{{S1}} with a counter equal to 2 (\includegraphics[scale=0.031,trim=0 40 0 0]{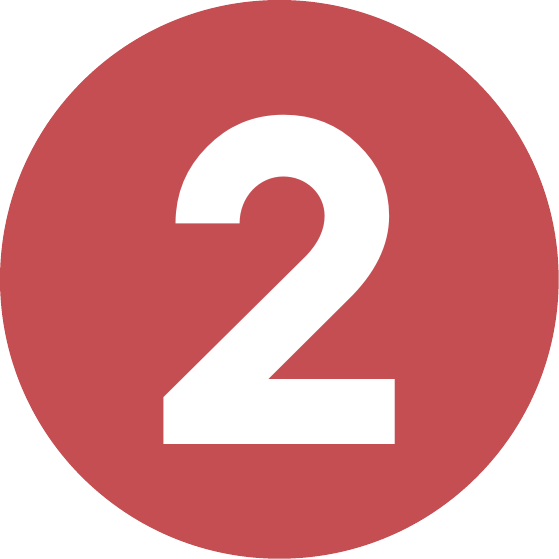}). At some point the accumulated counter will be equal to the number of hosts (\includegraphics[scale=0.031,trim=0 40 0 0]{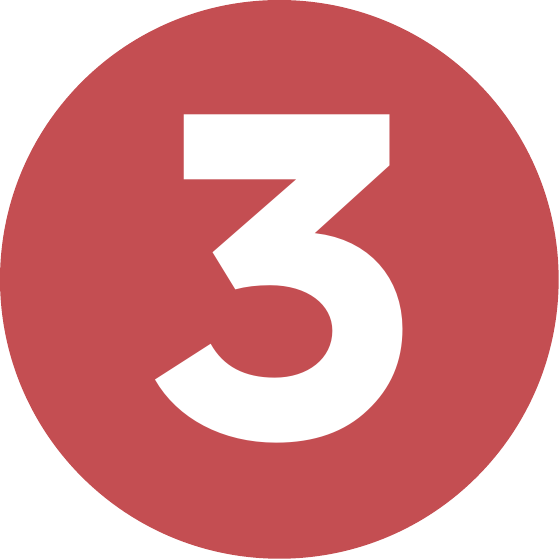}), meaning that all data coming from the hosts has been reduced and that the root can start the \broadcast phase. 

Intermediate switches might receive some packets after the timeout expiration if the timeout is too short. In that case, the packet is identified as a straggler and immediately forwarded to the next hop. In turn, the following switch considers that packet as a straggler or not, depending on its timeout. The switch can determine if a packet is a straggler because it does not deallocate the descriptor until the reduction is completed (which cannot happen unless all the packets are received and aggregated by the root). On the other hand, a too large timeout might increase the latency of the packets and the completion time of the allreduce. However, this is noticeable only for small allreduces, as we show in Section~\ref{sec:evaluation:sst:size}.

\subsubsection{Broadcast}
When the \broadcast starts, the root retrieves from the descriptor the list of the ports from which it received the data (i.e., the list of its children), forwards the aggregated data on those ports, and then deallocates the descriptor. After receiving a reduced packet from its parent, a switch forwards the packet to its children and deallocates the descriptor. In a nutshell, \canary reserves resources in the switches dynamically and strictly for the time required to complete the \reduce and \broadcast phases. A switch allocates a descriptor when it receives the first packet going to the root and deallocates it when it receives the packet coming down from the root. Eventually, the reduced data reaches the hosts that started the reduction, and the allreduce operation terminates. 

We can notice that, whereas \canary dynamically routes packets in the \reduce phase, in the \broadcast phase, it forwards them on the same paths used in the \reduce phase. A fully dynamic multicast would require explicit deallocation of resources because packets might cross different switches than those used in the \reduce phase. This would add unnecessary complexity to the design of the algorithm because, as we show in Section~\ref{sec:evaluation}, \canary can still bypass most of the congestion in the network. Also, as we discuss in Section~\ref{sec:design:general:packets}, the reduction tree is dynamically rebuilt on a packet-by-packet basis, thus reducing the probability of finding persistent congestion in the \reduce phase.

\begin{figure}[h]
    \centering
    \includegraphics[width=\columnwidth]{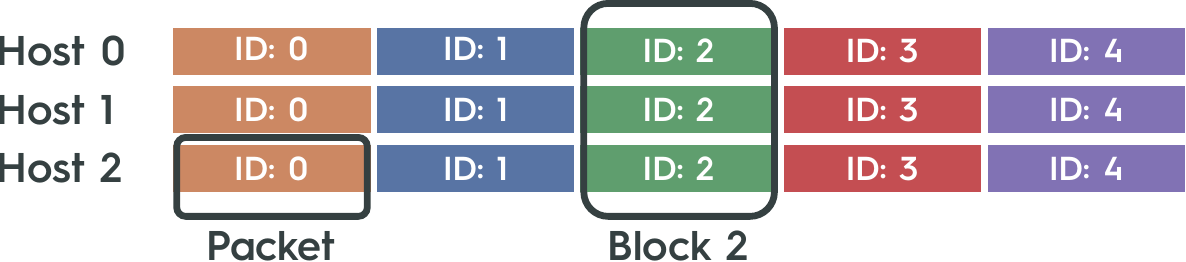}
    \caption{Packets and reduction blocks.}
    \label{fig:blocks}
\end{figure}

\subsubsection{Packetization}\label{sec:design:general:packets}

We assumed so far that all the data to be reduced by a host would fit in a single network packet. However, this is usually not the case, and data is often larger than the MTU (\textit{Maximum Transmission Unit}). Thus, each host assigns a unique identifier (\textit{id}) to each packet it sends, as shown in Figure~\ref{fig:blocks}. Packets with the same \textit{id} belong to the same \textit{reduction block} and must be aggregated together by the switch. The switches can now process multiple blocks concurrently and store a separate descriptor for each block in a table indexed by \textit{id}. Alongside the data accumulator, the list of children, and the timer, the descriptor also contains the block \textit{id}.

For the moment, we assume that there is always space available to store the descriptor. We then describe in Section~\ref{sec:design:collisions} how \canary works when the descriptor cannot be stored. To further improve load balancing, \canary aggregates each block in a different root, determined before starting the application (e.g., the hosts could select the roots in a round-robin way). 

\subsubsection{Leader Host}\label{sec:design:leader}
Programmable switches have limited resources and can only perform simple operations. As a consequence, it is not possible to fully implement complex tasks such as tracking and retransmission of lost packets or recovery from other switches failures, and \canary delegates these tasks to a \textit{leader host}, similarly to what happens in other algorithms~\cite{atp}. Reduction packets are still sent towards the root and aggregated on its path, as we described above. However, the root switch sends the aggregated data to the {leader host} as soon as the timeout expires (or when all the expected data is received). 

If we consider the example in Figure~\ref{fig:design_topology}, we could use switch \textit{S4} as root switch, and either \textit{H2} or \textit{H3} as leader host. The leader does not send its data on the network. Instead, it waits for data to arrive from \textit{S4}, after which it aggregates the received data with its own, and then starts the \broadcast phase. It is worth remarking that \canary still relies on a root switch that should aggregate as much data as possible. Ideally, the root should send to the leader host only one fully aggregated packet, thus avoiding having the leader receive multiple packets per block, which would reduce the operation's bandwidth. 

Although this solution increases the latency because packets need to cross the network stack of the leader~\cite{10.1145/3452296.3472888}, it allows us to partition tasks between hosts and switches. The switches only perform simple tasks, like aggregating data packets in a best-effort fashion. On the other hand, the {leader host} handles more complex operations like the retransmission of lost packets (Section~\ref{sec:design:losses}).
To reduce the latency required to cross the network stack, leader functionalities could be offloaded to programmable NICs~\cite{246498, pspin, inca, spin-transfers} or implemented on a high-performing network stack such as DPDK~\cite{dpdk}.
Moreover, because \canary sends each reduction block to a different leader, the pressure on individual hosts is reduced. Namely, suppose $N$ hosts participate in the reduction. In that case, each host will be the leader only for $1$ block out of $N$, thus receiving data at a much slower rate than line rate, increasing the time budget available for performing leader tasks.

\subsection{Resource Management}\label{sec:design:collisions}
Switches have limited memory resources and could not allocate memory for all the reductions concurrently executed over the network. Because \canary relies on adaptive routing, it cannot reserve any resource because the paths that packets will take are not known a priori, and reserving resources on all the network switches would unnecessarily increase resources occupancy. Instead, \canary stores block descriptors in a static array and, when a packet is received, the switch maps the \textit{id} to a specific array location (e.g., by using a hash function). If the location is empty, the switch stores data in the accumulator and initializes the descriptor. Otherwise, if the stored accumulator belongs to the same \textit{id}, the switch aggregates the data carried by the packet with that in the accumulator. If the stored accumulator belongs to a different \textit{id}, then we have a \textit{collision}. Collisions might happen because the hash function might map to the same location reduction blocks with different \textit{id}s (and that do not need to be aggregated together).

\subsubsection{Collisions Management}\label{sec:actualcollision}
\begin{figure}[h]
    \centering
    \includegraphics[width=\columnwidth]{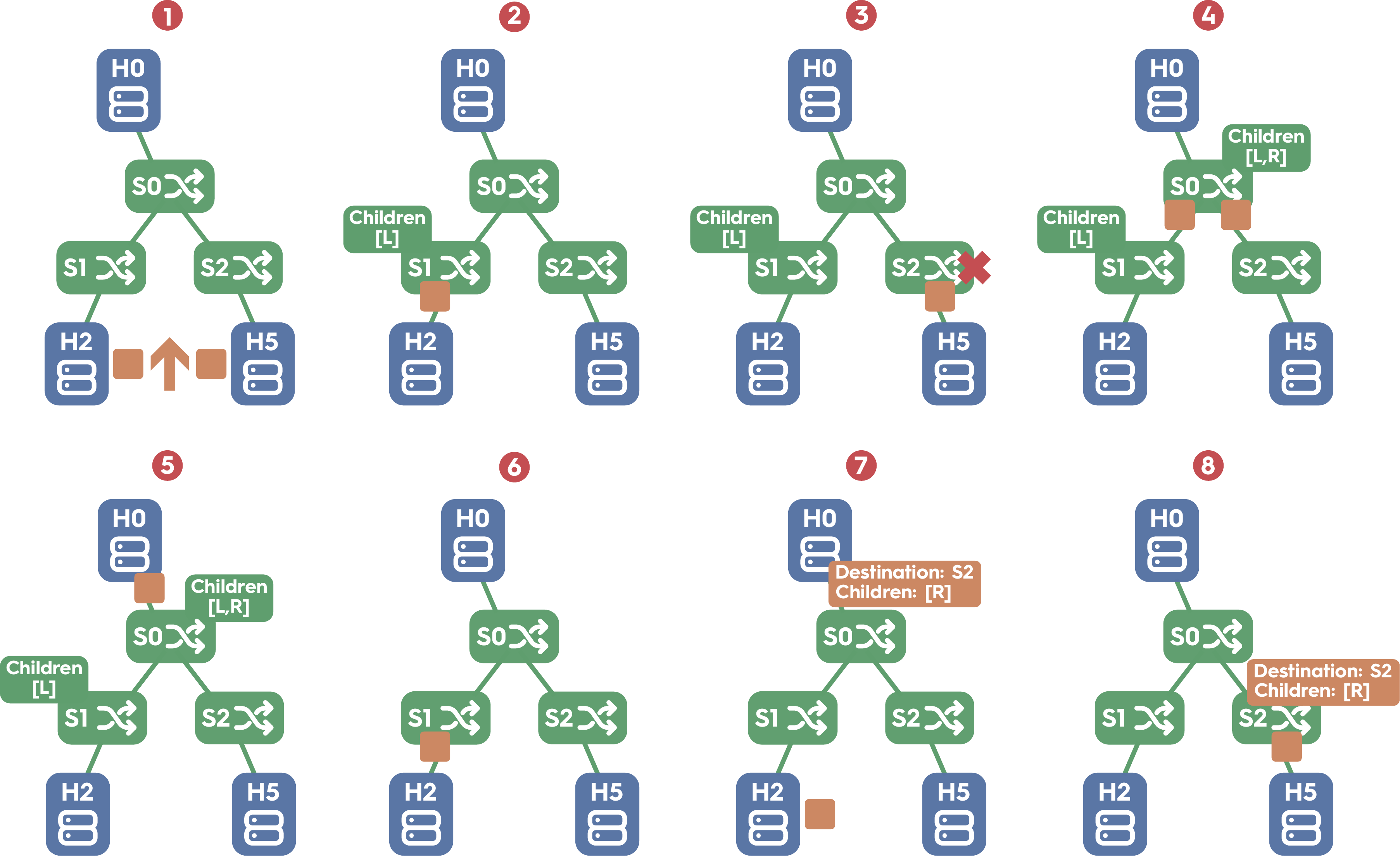}
    \caption{Collision and tree restoration.}
    \label{fig:design_collision}
\end{figure}

If there is a collision, the switch cannot store the descriptor of the new block. In principle, the switch could forward the packet to the next hop, delegating the aggregation to the following switches towards the root. However, in this case, the switch would not be able to participate in the subsequent \broadcast phase as it did not store the descriptor (containing the list of children) for that \emph{id}.
For example, in Figure~\ref{fig:design_collision} (\includegraphics[scale=0.031,trim=0 40 0 0]{marker_1.pdf}) hosts \textit{{H0}} (leader), \textit{{H2}}, and \textit{{H5}} want to reduce their data.
\textit{{S1}} receives data from \textit{{H2}} and experiences no collisions (\includegraphics[scale=0.031,trim=0 40 0 0]{marker_2.pdf}). 
\textit{{S2}} receives data from \textit{{H5}} (\includegraphics[scale=0.031,trim=0 40 0 0]{marker_3.pdf}) but detects a collision. Let us assume that \textit{{S2}} just forwards the data to the next hop and the reduction progresses
as normal from that point on.

During the \broadcast phase (\includegraphics[scale=0.031,trim=0 40 0 0]{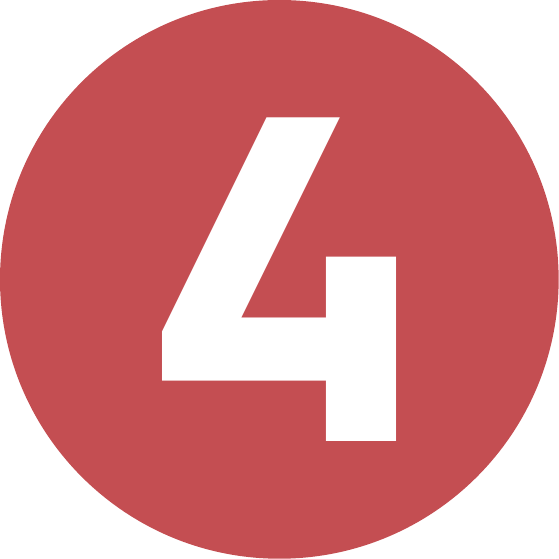}), \textit{{S2}} will not be able to forward data to its children because, due to the collision during the reduction phase, it was not able to store the descriptor that would contain, among others, the child port identifier. Ultimately, \textit{{H5}} will not receive the reduced data. In general, if a switch cannot store the block descriptor because of a collision, the entire subtree rooted at that switch will not be reachable during the \broadcast phase. It is worth remarking that each switch needs to store the ports connecting it to its children. Indeed, the leader host cannot just insert the addresses of the children of all the switches in the packet because this would be linear in the number of hosts participating in the reduction.

To solve this problem, we adopt a simple but effective solution called \emph{tree restoration}. After a conflict, the switch inserts its address in the packet alongside the identifier of the port from which it received it. Then, it forwards the packet directly to the leader host (the packet is marked and ignored by all the other switches on the path). After the reduction, the leader host knows the unreachable subtrees: i.e., it has a list of switches and respective ports from which they received packets generating collisions. During the \broadcast phase, the leader host uses this information to send an additional packet to these switches, allowing them to bootstrap a local broadcast, restoring the subtree. Other than the reduced data, these packets also carry the list of ports on which the switch must forward the data (e.g., encoded as a bitmap). 

In our example, when the collision is detected (\includegraphics[scale=0.031,trim=0 40 0 0]{marker_3.pdf}), \textit{{S2}} forwards the packet to the leader host, together with its address and the port number (\textit{R}) from which the packet has been received. After the reduction phase, the leader host \textit{{H0}} starts the \broadcast as usual (\includegraphics[scale=0.031,trim=0 40 0 0]{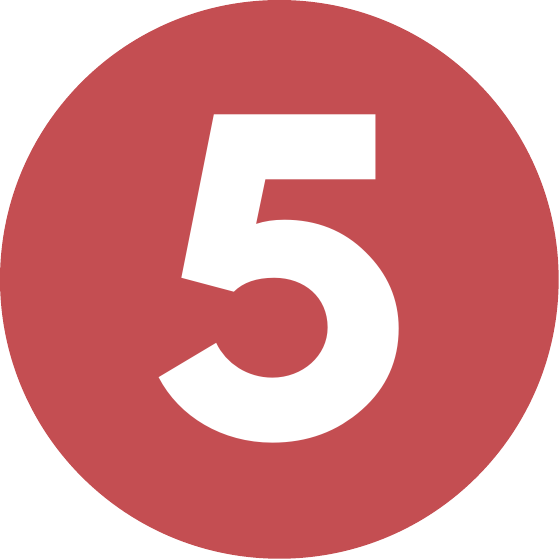}). The broadcast packet eventually reaches \textit{{S2}}, but it is not able to progress since \textit{{S2}} is missing information about its subtree (\includegraphics[scale=0.031,trim=0 40 0 0]{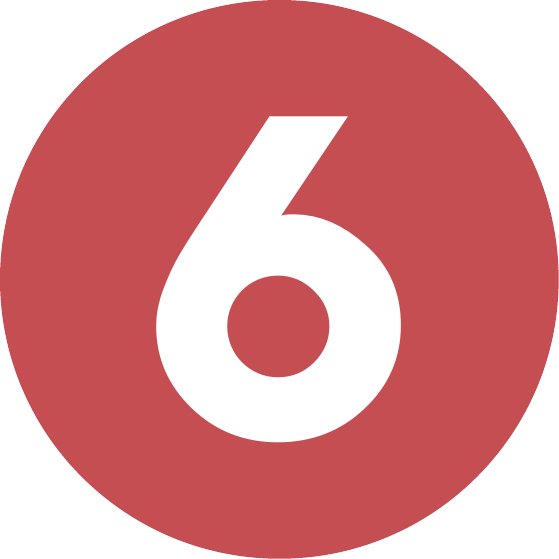}). The leader host also sends a \textit{``restoration''} packet to \textit{{S2}} (\includegraphics[scale=0.031,trim=0 40 0 0]{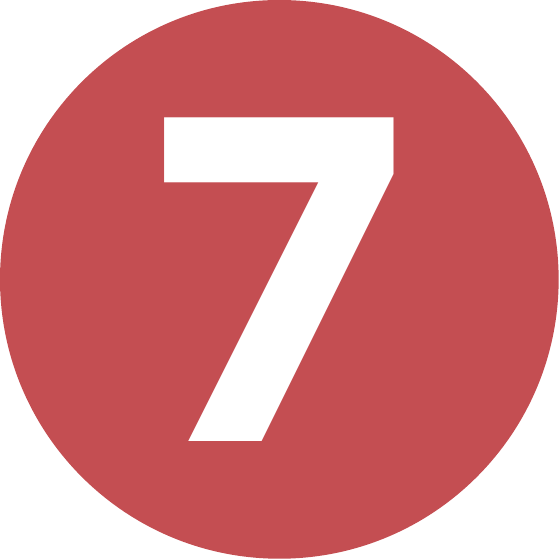}), making it able to forward the data on port \textit{R} towards \textit{{H5}}  (\includegraphics[scale=0.031,trim=0 40 0 0]{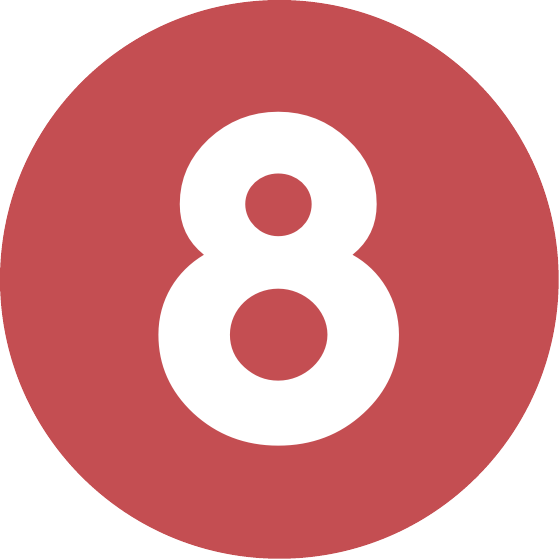}).

This approach works even if, after some collisions, the entry becomes available and subsequent packets are successfully stored. In that case, the switch stores identifiers of some children in the block descriptor on the switch, and hence they will be reached by the normal \broadcast phase. Others, i.e., the ones that generated conflicts, will be reached through the tree restoration process. 

Because of the extra network traffic generated after a collision (e.g., restoration packets), this approach can lower the throughput (i.e., the leader host receives more packets due to missed aggregations). However, collisions only happen if a switch receives multiple packets with different \textit{id}s mapping to the same descriptor entry in the same time window. If this performance penalty is not acceptable, \canary can avoid collisions entirely by setting a limit on the number of concurrent allreduces, and by statically mapping \textit{id}s to descriptor array entries. 

\subsubsection{Switch Memory Occupancy Modelling}\label{sec:design:resource:occupancy}
We now analyze how much switch memory an allreduce can occupy. A block descriptor is allocated in a switch when the first packet of that block is received and deallocated when the fully aggregated data is received. For this reason, switches at the bottom level of the tree keep descriptors allocated for the longest time and, to estimate the maximum memory occupancy, we model the memory occupancy of those switches. 

We denote the network bandwidth with $b$, the network diameter with $d$, the 1-hop delay with $l$, the timeout with $t$ (i.e., how much time a switch waits before sending the partially aggregated data to the next hop), and the time required to the leader to aggregation its data with $r$. Then, the time between the allocation and the deallocation of a descriptor can be measured as $2 d (l + t) + r$. Each descriptor contains the aggregated data, plus a few more bytes for storing the root address and other information. Thus, we can approximate the size of a descriptor with the MTU $m$. By using Little's Law, if we assume to send MTU-sized packets, we can estimate the number of bytes occupied by descriptors as:
\[
\frac{b}{m}  (2  d  (l+t) + r)  m = b (2  d  (l+t) + r)
\]

Recent networks have a diameter of up to five and a per-hop latency of around 300 nanoseconds~\cite{slingshot}, and programmable NICs can perform tasks similar to those performed by the leader host in around one microsecond~\cite{pspin}. Thus, on a 100Gbps network with a one-microsecond allreduce timeout, each allreduce might store up to 175KiB in each switch crossed by its packets. 

It is worth remarking that the memory occupancy is independent of the actual size of the data to be reduced because \canary aggregates data block-by-block, and the bandwidth-delay product bounds the number of in-flight blocks. Also, the occupancy does not depend on the number of hosts participating in the reduction. Indeed, each switch stores only one descriptor for each block, independently of how many packets (one from each child) are aggregated in that block.

\subsection{Packet Loss and Fault Tolerance}\label{sec:design:losses}
\canary treats packet losses and switches failures in the same way. Indeed, in both cases, the leader does not receive some packets (if the loss/failure occurs in the reduce phase), or the hosts do not receive the reduced data (if the loss/failure occurs in the broadcast phase). Without loss of generality, we describe how to manage packet losses since the same approach is also used for managing switches failures. 

To detect a loss, all the hosts (excluding the leader) set a timeout for each packet right before transmitting it. When the reduced data arrives, the host deletes the timer. If the timeout expires, a retransmission request is issued. If the leader receives a retransmission request, two situations might occur. If the leader entirely reduced the data, the packet was lost during the broadcast phase, and the reduced data is re-transmitted to the host that issued the request. Otherwise, if the leader only partially reduced the data, some packet was lost in the reduce phase. 

In this case, the leader does not know which packet was lost. Indeed, to know which packets contributed to the partially reduced data, each switch would need to keep a bitmap of all the hosts that contributed to the data reduced so far, which would be linear in the number of hosts participating in the reduction. However, having this bitmap in the packet is infeasible because allreduces might span thousands of hosts~\cite{ddl_survey,10.1109/SC.2018.00033}, and existing programmable switches can only process a few hundred bytes per packet (Section~\ref{sec:discussion}). Accordingly, because the leader cannot determine which packet should be re-transmitted, it broadcasts a failure message. Upon the reception of this message, the hosts re-issue the reduction of that packet with a different \textit{id} (or they can reduce that packet only by using a host-based reduction algorithm). To avoid overloading the network with reduction packets, the hosts fall back to a host-based reduction after a given number of failed retransmissions. %

It is worth remarking that when a host terminates the reduction, it cannot simply modify or deallocate the reduced data because the other hosts might not have successfully terminated the reduction yet. Accordingly, it must preserve the part of the data for which it was the leader to re-transmit the packets in case any retransmission request arrives. For small-size reductions, the leader can store a copy of the data and deallocate it when it receives a fully reduced packet for a subsequent reduction. Indeed, the hosts can start the subsequent reduction only if they have already completed the previous one. If there are no subsequent reductions, an explicit completion notification is required (for example, by issuing a barrier). Because preserving the data between subsequent reductions could potentially double the memory consumption, for large reductions the hosts always explicitly notify the completion. The explicit notification introduces a marginal latency overhead compared to the allreduce and allows the data to be deallocated or modified immediately after the notification, not requiring any additional copy. 

Although \canary can autonomously manage switch failures without re-issuing the entire allreduce operation, host failures must be managed at a higher layer (e.g., with checkpoint/re-start solutions).

\subsection{Multitenancy}\label{sec:design:multitenancy}
The switch does not have any knowledge of the different applications or users running on the system and simply aggregates together packets with the same \textit{id}. To support multiple applications, each of them must generate unique \textit{id}s. Thus, \canary \textit{id}s are built by concatenating an identifier of the application (e.g., generated by the workload manager) and an identifier that each application increases for every subsequent packet. 

It is worth remarking that having multiple concurrent allreduces does not necessarily increase the amount of data that a switch needs to store. Indeed, a descriptor is allocated only on the switches traversed by the packets belonging to the corresponding block and strictly for the time needed to reduce that block. In a nutshell, running multiple concurrent allreduces, each on a few hosts (thus connected through a few switches), might consume the same amount of resources of a single allreduce on a higher number of hosts (thus spanning over more switches). %

\subsection{Summary}
We now wrap up \canary design. First, each host splits the data in multiple packets, each marked with a unique \textit{id}. When a switch receives a data packet, it maps the \textit{id} to a specific entry of the array containing the descriptor for that \textit{id}. If the entry is available, the switch stores in the table the data carried by the packet, updates the list of children, starts a timer, and drops the packet. If a descriptor with the same \textit{id} is already present in the entry, the switch accumulates the data, updates the list of children, and drops the packet. If a descriptor with a different \textit{id} already occupies the entry, the switch inserts in the packet its address and the identifier of the port from which it received the packet, before forwarding it to the leader.

When the timer of a descriptor expires, the switch sends to the next hop the data contained in the descriptor. The port is selected based on the root address stored in the switch using any available load balancing algorithm. If a packet arrives and the timeout for that \textit{id} already expired, the switch updates the list of children and forwards the packet to the next hop. Eventually, when the leader starts the \broadcast phase, the switch receives the fully reduced data. 

If there is a descriptor for that \textit{id}, the switch forwards the packet to all its children and removes the descriptor from the table. If the switch does not have a descriptor for that \textit{id} (because it could not store it due to a collision), it drops the packet. Later, it receives the data from the leader specifying the children to which the packet should be forwarded. 

At some point, all the data will reach the tree's leaves. If a packet is lost or a switch fails, some leaves send a retransmission request to the leader. When the leader receives a retransmission request, it can either re-transmit the fully reduced data or require that block to be reduced from scratch (if it did not already entirely reduce the data).

\section{Implementation}\label{sec:implementation}
We implemented \canary using a state-of-the-art Intel Tofino programmable switch. Although existing programmable switches can process terabits of data per second (1.2 billion packets per second for each pipeline), they limit the type of computation they can perform on packets. These limitations drove some of the main choices in our design. For example, having one of the hosts acting as the reduction leader pushes some complexity to the host and keeps the code executed by the switch as simple as possible. 

Programmable switches process packets through multiple pipelines, each composed of multiple stages. In our implementation, we use the first stages to determine if the packet is a reduce or a broadcast packet and check if the packet generates any collision when accessing the descriptors table. We then use subsequent stages to read/write data. Because the goal of \canary is to leverage the speed of such programmable devices, all the data is managed in the \textit{data plane} and stored into registers. Content-addressable memory (CAM)~\cite{cam} is usually available but can only be updated from the \textit{control plane}~\cite{sdn,tofino}. We do not use CAM for storing the data because interactions with the control plane would significantly increase the latency~\cite{controlplane_lantency}.

\subsection{Packet Format}
Because we are using programmable switches, we define a custom packet format for \canary packets. To reduce the packet overhead, \canary sends packets directly on top of Ethernet. However, any other encapsulation could be used, and \canary packets could be sent on top of IP or UDP. A \canary packet is composed of the following fields:
\begin{itemize}[leftmargin=*]
\item \textbf{Destination (4 bytes)} IP address of the leader host. \canary uses the same routing tables used for IP routing to determine how to reach the destination.
\item \textbf{Id (4 bytes)} Unique identifier of the packet. 
\item \textbf{Counter (2 bytes)} Number of reduced packets (Fig.~\ref{fig:design_topology}). 
\item \textbf{Hosts (2 bytes)} Number of hosts participating in the reduction (Fig.~\ref{fig:design_topology}).
\item \textbf{Children (4 bytes)} When a switch cannot store a packet because of a collision, this field carries the identifier of the port from which the packet was received (Section~\ref{sec:design:collisions}).
\item \textbf{Switch Address (2 bytes)} When a switch cannot store a packet because of a collision, this field carries the switch address (Section~\ref{sec:design:collisions}).
\item \textbf{Bypass (1 bit)} If set, the switch should not further process the packet but only forward it to the next hop.
\item \textbf{Multicast (1 bit)} If set, the packet must be multicast to the children of the switch. 
\item \textbf{Padding (6 bits)} Used to pad the packet size to a multiple of 8 bits. 
\item \textbf{Data (128 bytes)} Data to be reduced. 
\end{itemize}

\subsection{Multicasting}\label{sec:implementation:multicast}
Multicasts could, in principle, have an impact on the capacity to run our allreduce algorithm at line rate. Indeed, if the switch generates $m$ packets for each packet it receives, it might decrease the achievable bandwidth by a factor of $m$. However, we observe that if a switch multicasts a packet on $m$ ports (its children), this is because it previously aggregated the data coming from $m$ ports. Accordingly, the switch forwards, on average, one packet for each packet it receives. 

The switch keeps a table associating to each port the corresponding one-hot encoding. Every time a packet is received, the one-hot encoding of the input port is retrieved through a TCAM (pre-configured in the control plane) and added to the bitmap storing the children. Thus, when a packet needs to be multicasted, the switch knows that it must send it to all the ports set in the children bitmap.

Programmable switches require multicast groups to be pre-configured by specifying the association between a group identifier and a list of ports on which the switch will send packets directed to that group. In our case, the group identifier could simply be the bitmap associated with the specific list of ports. For example, let us suppose we have a switch with eight ports and that we want to multicast on the ports \texttt{[0,2,3,5]}. The binary representation of this list is \texttt{00101101}. So we would have to set up a rule such as \texttt{00101101 -> [0,2,3,5]}. However, \canary uses adaptive routing, and the switches multicast the packets to the same ports from which they have been received, which are not known a priori. Accordingly, we should store all the possible combinations of ports, which is exponential in the number of ports. 

Because this requires too many resources, to reduce the storage requirements, we divide the children bitmap in \textit{shards}. For example, the children bitmap \texttt{00101101} can be divided into two shards of 4 bits each. We prepend to each shard its index so that the two shards become \texttt{1 0010} and \texttt{0 1101}. We then store the association between all the possible shard values (that, in this case, would be $2\cdot2^4$) and the corresponding list of ports. In our example, we would have the rules \texttt{10010 -> [5]} and \texttt{01101 -> [0,2,3]}. This technique reduces the number of multicast groups to store in the switch tables from $2^p$, to $2^\frac{p}{s} \cdot s$, where $p$ is the number of ports, and $s$ is the number of shards. For example, on a 64 port switch with four shards, this requires using 256 thousand entries, which is far within reach of current programmable switches~\cite{201474}, as we also show in Section~\ref{sec:evaluation:switch}.

\subsection{Timeouts}\label{sec:implementation:timeout}
As described in Section~\ref{sec:design:general:reduce}, \canary relies on timeouts to avoid statically setting up the reduction tree. In our \canary implementation, every time the switch receives a reduction \textit{id} for the first time, it stores in the descriptor the current timestamp (alongside the data to reduce and the other information). Some programmable switches~\cite{tofino} provide one or multiple \textit{packet generators} that can generate packets at a predefined rate and with predefined content. For example, we can set up the packet generators so to generate \textit{clock} packets. Every time one of these packets is received, the switch can check an entry of the table and, if expired, send the content in that entry to its parent.
Alternatively, \canary could be implemented on programmable switches that support timer-based events~\cite{pisa-event}.

\section{Evaluation}\label{sec:evaluation}
In this section, we evaluate the performance of \canary, first by analyzing its performance on a Tofino switch (Section~\ref{sec:evaluation:switch}), and then by simulating it on a large network (Section~\ref{sec:evaluation:sst}) connecting 1024 hosts through 64 switches.

\subsection{Single switch implementation}\label{sec:evaluation:switch}
To verify the feasibility of our design, we implemented and validated our P4 \canary prototype on a Tofino Wedge 100BF-32X switch~\cite{tofino} with 32 100Gbps ports. This allowed us to understand existing switches' limitations and drive our design choices. We were able to allocate enough registers to allocate up to 32K descriptors, allowing us to run at least 25 concurrent allreduces of different tenants or applications (Section~\ref{sec:design:resource:occupancy}).
Our P4 \canary implementation only uses 14.38\% of the switch SRAM. However, \canary uses most of the Arithmetic Logic Units (ALU), up to 81.25\% of those available on the switch, to aggregate the elements carried in the packets. 
\canary leaves enough resources available for running it alongside traffic load balancing algorithms such as flowlet switching~\cite{conga} that, on the same switch, uses 2.26\% of the available SRAM and 0.04\% of the ALUs~\cite{electronics8090958}.

\begin{figure}[h]
    \centering
    \includegraphics[width=\columnwidth]{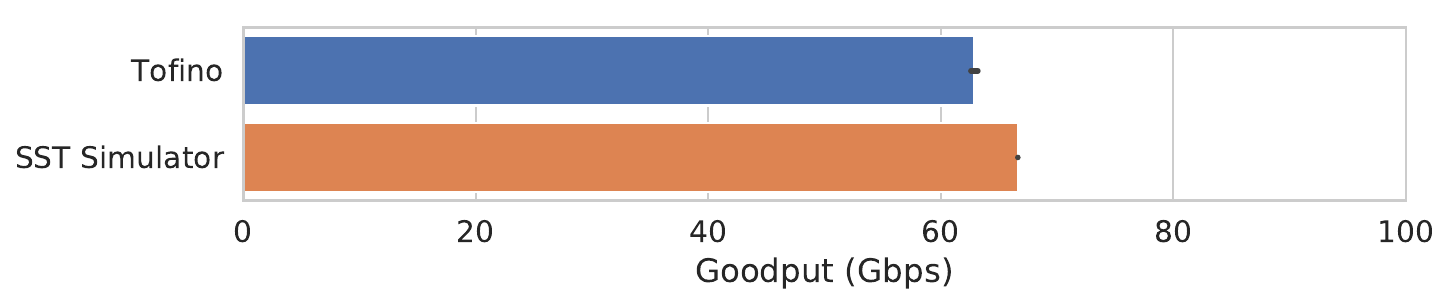}
    \caption{Goodput (Gbps) of our P4 \canary prototype and of our implementation in the SST simulator, when sending packets with 128 bytes of useful payload.}
    \label{fig:results:tofino}
\end{figure}

To measure the goodput of our P4 prototype implementation, we connected two hosts equipped with 100 Gbps Mellanox ConnectX-5 NICs to the Tofino switch. We emulate a leaf switch that receives the data to be reduced from the two hosts, aggregates it, and forwards the aggregated result to the next switch in the reduction tree. The two hosts inject the data using Moongen~\cite{moongen}, a DPDK~\cite{dpdk} wrapper. 

We benchmark a 4MiB allreduce and report the goodput in Figure~\ref{fig:results:tofino}. We also report the goodput achieved by our simulation infrastructure (that we describe in Section~\ref{sec:evaluation:sst}) in the same setup. It is worth remarking that, due to the limited number of \textit{match-action} tables available in existing programmable switches, we can store up to 32 4-bytes elements. Accordingly, each packet contains 128 bytes of useful payload and 57 bytes of headers (19 bytes of \canary header, 14 bytes of Ethernet header, and 24 bytes of framing overhead).

Programmable switches are composed of multiple processing pipelines, and some existing P4 prototypes~\cite{atp,switchml} partially overcome this limitation by striping the packet across pipelines. For example, the first pipeline would store the first 32 elements, then recirculate the packet to the second pipeline, which would store the following 32 elements, and so on. In this way it is possible to have up to 128~\cite{atp} or 256 elements~\cite{switchml} per packet. However, to have enough recirculation bandwidth and avoid packet drops, some switch ports must be dedicated to packets recirculations only. 

Additionally, it is reasonable to assume that the number of match-action tables in each pipeline stage will increase with the next generations of programmable switches, thus allowing processing more elements per packet. For example, some programmable switches already provide the possibility to process up to 48 elements per pipeline~\cite{electronics8090958}. For these reasons, in the following, we run simulations with 256 elements per packet for all the in-network algorithms.

\subsection{Large network simulations}\label{sec:evaluation:sst}
To evaluate \canary performance at scale, we modified the SST simulator~\cite{sst-0,sst} so that switches can modify the packets they receive before forwarding them. We build in the simulator a two-level fat-tree network~\cite{fattree}. The network comprises 32 switches at the bottom level, each with 64 ports (32 connected to the hosts and 32 to the switches at the upper level). The top level of the fat-tree comprises 32 switches, each with 32 ports (one port connected to each of the switches at the bottom level). Both the hosts and the switches have 100 Gbps network interfaces.

We calibrated the simulator so that hosts can inject packets into the network at line rate and so that the switches can aggregate the data at the same speed as our Tofino prototype (as we show in Figure~\ref{fig:results:tofino}). The simulated network uses up/down routing. When packets flow from hosts to upper levels of the fat-tree, each switch can select one among multiple \textit{up} ports. By default, each switch sends packets on a default \textit{up} port (selected depending on the packet destination). If the output port buffer has an occupancy higher than 50\% of its capacity, the switch forwards the packet on the \textit{up} port with the smallest number of enqueued bytes.

To compare our solution with the state-of-the-art, we implemented in the simulator the following allreduce algorithms:
\begin{itemize}[leftmargin=*]
\item \textbf{Ring} The bandwidth-optimal host-based \textit{ring allreduce} algorithm~\cite{bwoptimalallreduce}. This solution does not rely on any in-network compute capability.
\item \textbf{In-Network, $N$ static trees} An in-network algorithm using static reduction trees. We consider either the case when a single tree is used, similar to what is done by SHARP~\cite{sharp,sharp2}, SwitchML~\cite{switchml}, and ATP~\cite{atp}, and also the case where $N$ trees are used and each block is sent on a different tree in a round-robin way, similar to what done by PANAMA~\cite{panama}.
\item \textbf{\canary} The in-network algorithm we propose in this work, which dynamically builds reduction trees.
\end{itemize}

To analyze the impact of congestion on the allreduce performance, we split the hosts into two sets. While some hosts run the allreduce, the remaining hosts generate network congestion by executing a random uniform injection traffic pattern. In this pattern, each host sends a message to a randomly selected host and receives a message from another randomly selected host. Each host changes its random peer throughout the execution to assess the ability of \canary to react to dynamically changing congestion patterns. We execute each test five times, each time randomly allocating the hosts executing the allreduce and those generating the congestion. When using static reduction trees, we also randomly pick the roots of those trees.

\begin{figure}[htpb]
\begin{subfigure}[b]{.5\columnwidth}
  \centering
    \includegraphics[width=\linewidth]{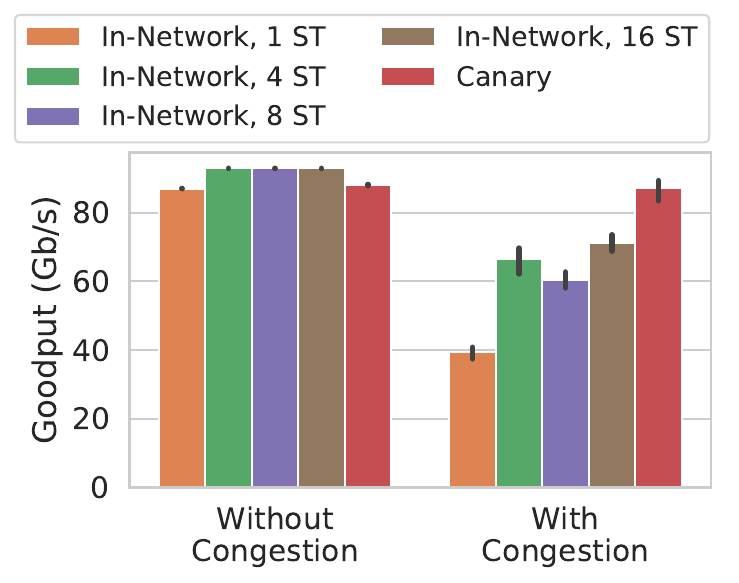}
    \caption{Goodput (Gbps).}
    \label{fig:experiments:multi_trees_bandwidth}
\end{subfigure}%
\begin{subfigure}[b]{.5\columnwidth}
  \centering
    \includegraphics[width=\linewidth]{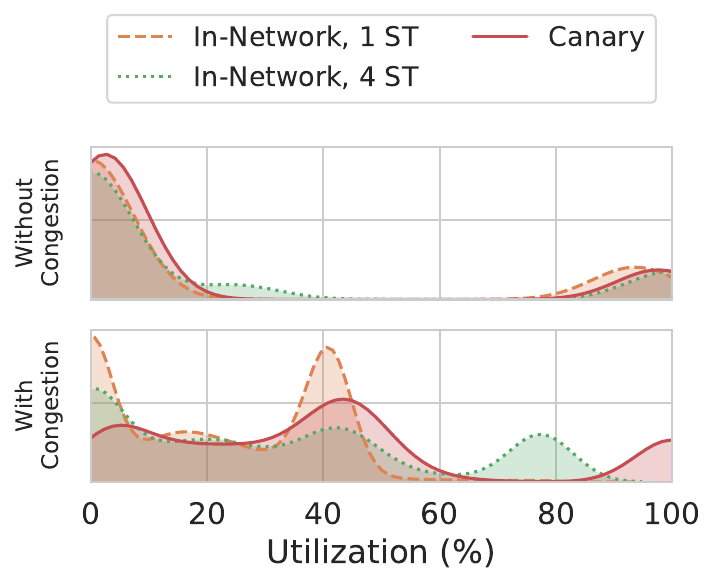}
    \caption{Links utilization (\%).}
    \label{fig:experiments:multi_trees_utilization}
\end{subfigure}
\caption{Goodput and links utilization when 512 hosts execute an allreduce and 512 hosts generate congestion. \textit{ST:} Static Tree(s).}
\label{fig:experiments:multi_trees}
\end{figure}

\begin{figure*}[h]
    \centering
    \includegraphics[width=\linewidth]{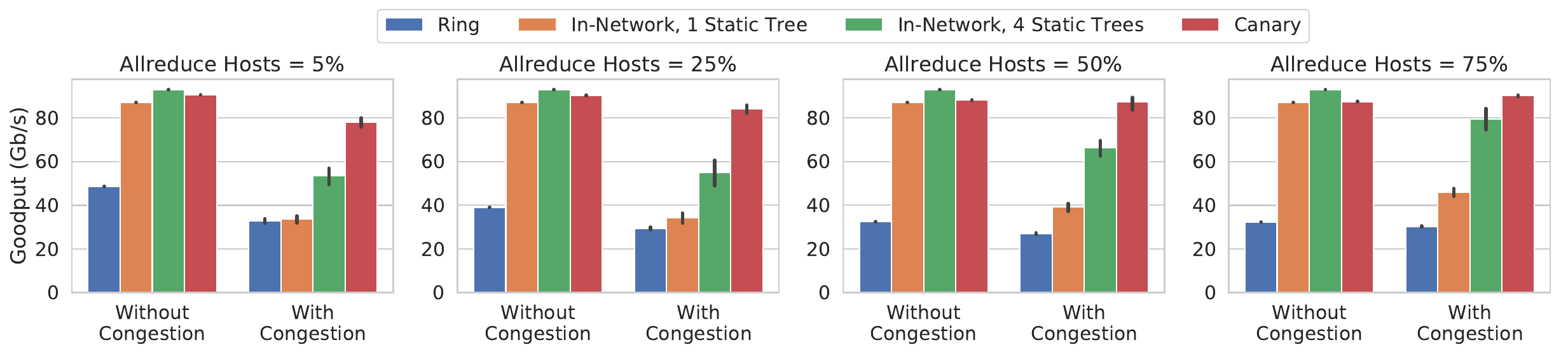}
    \caption{4MiB allreduce goodput (the higher the better) for different hosts count. The hosts not performing allreduce generate random uniform traffic to introduce congestion.}
    \label{fig:experiments:ratio}
\end{figure*}

\subsubsection{Comparison with the static trees approach} 
First, we report in Figure~\ref{fig:experiments:multi_trees_bandwidth} the comparison between \canary and the in-network algorithm using one or multiple static trees. We allocated 512 hosts to the allreduce and used the remaining 512 hosts to generate congestion. We report the bandwidth with and without congestion for a 4MiB allreduce. Whereas in the absence of congestion, the performance of all the approaches is comparable, when introducing congestion \canary performs significantly better than the solutions using statically configured reduction trees. Indeed, whereas solutions using static trees are severely affected by congestion, \canary does not experience any performance degradation. For this reason, we observe performance improvements up to 2x compared to solutions using a single reduction tree and up to 40\% compared to those using multiple trees. Moreover, we also observe that using more than four trees for solutions relying on static trees leads only to a marginal performance improvement. For these reasons, in the subsequent analysis, we consider a solution using four static trees, as in the original PANAMA paper~\cite{panama}.

We also report in Figure~\ref{fig:experiments:multi_trees_utilization} the distribution of links utilization (each sample is a network link). For the sake of readability, we only report those of \canary, one static tree, and four static trees. We observe that when there is no congestion, there are no significant differences between the three approaches, and each link is either idle (0\% utilization) or fully utilized (around 90\% utilization). However, when we introduce congestion, we observe that \canary is characterized by fewer idle links and better distributes the traffic over the available links. 

At first sight, it might seem like the in-network solution with one static tree does not fully utilize any network link because there are no humps around 80-100\% utilization. However, by analyzing the data more in detail, we found two links with utilization greater than 80\%. Because these two links are shared between the in-network allreduce and the application that generates congestion, this is enough to slow down the in-network allreduce by more than 50\% (Figure~\ref{fig:experiments:multi_trees_bandwidth}) and to reduce the overall network utilization. Indeed, in the presence of congestion, we observed an average network utilization (computed as the average of all the links utilization) of 40.2\% for \canary, 29.5\% for the in-network allreduce with four trees, and 20.9\% for the one with one tree.

\begin{figure*}[h]
    \centering
    \includegraphics[width=\linewidth]{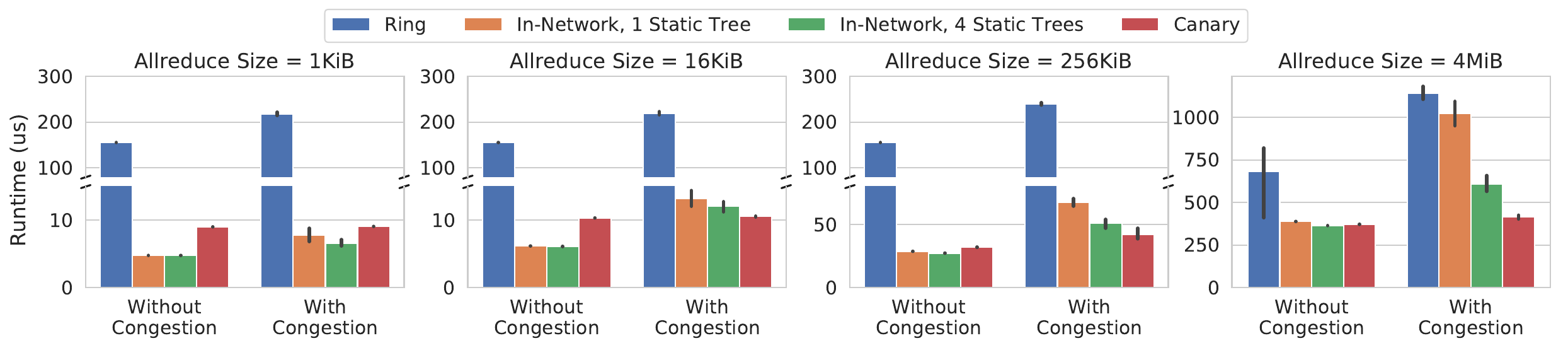}
    \caption{Allreduce runtime (the lower the better) for different message size, when $20\%$ of the hosts are allocated to the allreduce, and $80\%$ to an application generating random uniform traffic.}
    \label{fig:experiments:size}
\end{figure*}
\subsubsection{Goodput for different congestion intensity}
We now analyze the performance of \canary when changing the number of hosts generating congestion by comparing it to the host-based bandwidth-optimal ring allreduce and the in-network allreduce using static trees. We report the results of this analysis in Figure~\ref{fig:experiments:ratio}. We ran a 4MiB allreduce, and we increase the number of hosts executing the allreduce from 5\% to 75\% of the 1024 hosts available in the system. The hosts not executing the allreduce generate congestion through a random uniform communication pattern. First, we observe that \canary consistently improves performance compared to other solutions. 

When using only 5\% of the hosts for the allreduce (thus using 95\% of the hosts to generate congestion), \canary performance only decreases by 20\%. In contrast, the performance of the in-network static solutions decreases by 66\% when using a single tree and by 47\% when using four trees. When increasing the number of hosts executing the allreduce (thus decreasing congestion), the performance gap shrinks, but \canary still provides 2x improvement compared to the single static tree solution, and 23\% improvement compared to the solution using four reduction trees. Eventually, we observe that in some cases, the congestion decreases the performance of the single tree solution so much that it does not provide any performance advantage compared to the host-based ring allreduce, as also outlined in other recent works~\cite{panama}.

\subsubsection{Runtime for different data sizes}\label{sec:evaluation:sst:size}
We now analyze the allreduce runtime for different data sizes. We allocate $20\%$ of the hosts to the allreduce, whereas the remaining $80\%$ generates congestion.  We report in Figure~\ref{fig:experiments:size} the runtime (in microseconds) with and without congestion. We observe that for small allreduces, \canary is characterized by a higher runtime because the switch only forwards (aggregated) packets after the timeout period expires. When increasing the size of the data exchanged by the allreduce, the performance advantage of \canary increases because the runtime of large allreduces is dominated by the bandwidth, and the extra latency introduced by the timeout mechanism becomes negligible. We also observe that 1KiB and 256KiB ring allreduces have the same runtime. Indeed, the ring allreduce is the host-based bandwidth-optimal allreduce algorithm, but, for small messages, its runtime is dominated by latency and setup of communication phases~\cite{bwoptimalallreduce,nvidiaallreduce}.

\subsubsection{Multiple concurrent \textit{allreduces}}\label{sec:eval:concurrent}
Using statically configured trees also significantly decreases the aggregation bandwidth when multiple tenants (or multiple applications) concurrently run allreduce operations (e.g., multiple training jobs). Therefore, we equally partitioned the system between multiple co-running allreduce operations to analyze this effect. Furthermore, because most existing in-network allreduce algorithms statically partition the switch resources across the tenants~\cite{switchml,flare,sharp}, to have a fair comparison, we adopt a similar approach also in \canary.

\begin{figure}[htpb]
\begin{subfigure}[b]{.535\columnwidth}
  \centering
    \includegraphics[width=\linewidth]{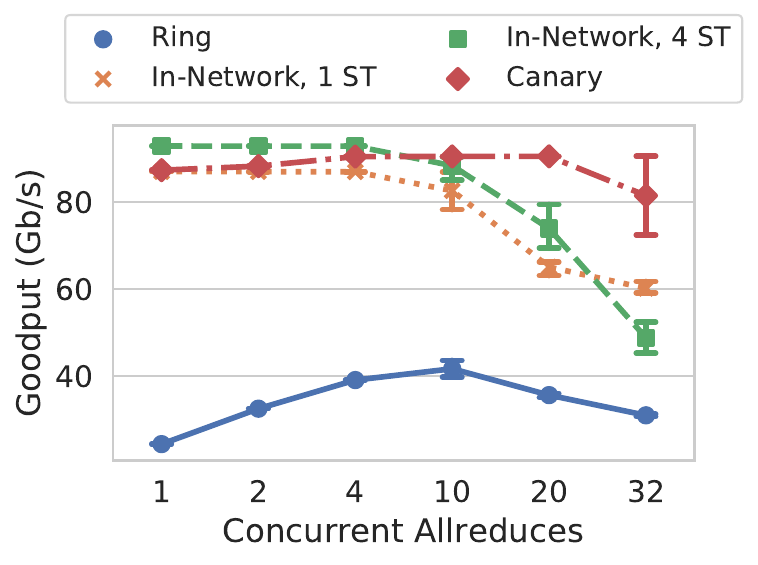}
    \caption{Goodput (Gbps).}
    \label{fig:experiments:concurrent_bw}
\end{subfigure}%
\begin{subfigure}[b]{.46\columnwidth}
  \centering
    \includegraphics[width=\linewidth]{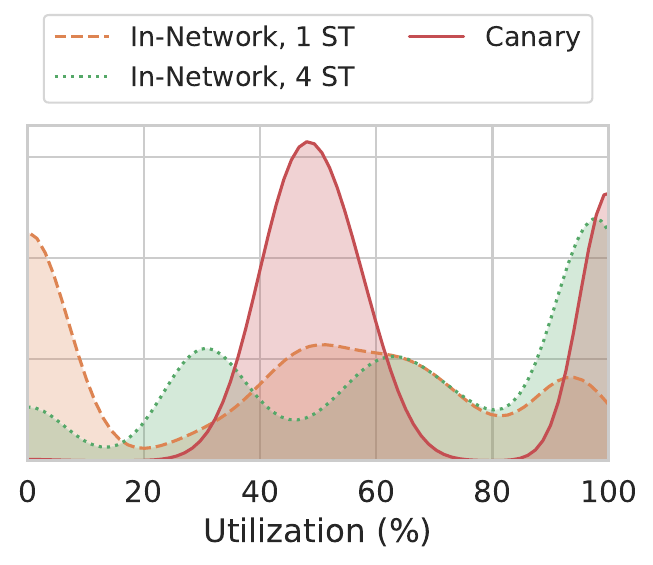}
    \caption{Links utilization (\%).}
    \label{fig:experiments:concurrent_util}
\end{subfigure}
\caption{Average goodput of multiple concurrent 4MiB allreduces (left), and link utilization when running 20 concurrent allreduces (right). \textit{ST:} Static Tree(s).}
\label{fig:experiments:concurrent}
\end{figure}

We report in Figure~\ref{fig:experiments:concurrent_bw} the average goodput across all the concurrent allreduce operations. First, we observe that when increasing the number of concurrent allreduces (thus decreasing the number of hosts allocated to each allreduce), the average goodput of the ring allreduce increases. This is a known effect~\cite{bwoptimalallreduce} because the performance of the ring allreduce increases when decreasing the number of hosts. However, the performance drops when running more than ten concurrent allreduces due to the increased congestion. We also observe the performance of in-network static allreduce algorithms drops by 40\% when increasing the number of concurrent allreduce operations due to congestion. On the contrary, \canary is almost unaffected and allows up to 32 concurrent allreduces at 80 Gbps each. %

By analyzing the distribution of the links utilization in Figure~\ref{fig:experiments:concurrent_util}, we observe that \canary is characterized by the lowest number of idle links. We also observe that using four static trees improves the links utilization compared to using only one static tree. However, as shown in Figure~\ref{fig:experiments:concurrent_bw}, this is still not sufficient to avoid congestion because multiple allreduces concurrently use some links. 

We observed an average network utilization of 67.2\% for \canary, 62.9\% for the static in-network allreduce with four trees, and 21.8\% for the one with one tree. Although \canary and the solution using four trees lead to a similar average network utilization, \canary performs better because it distributes the traffic more evenly across the network (e.g., it does not have any link in the 70-90\% range of utilization).

\subsubsection{\revb{Impact of Timeout and Noise}}
\revb{
One of the key points of \canary is the use of timeouts to decide when a reduction block has been fully aggregated and can be sent to the next hop in the reduction tree. As described in Section~\ref{sec:design:general:reduce}, if the timeout is too short, or if a packet is delayed for any reason (e.g. OS noise~\cite{10.1109/SC.2010.12}), the straggler packet is sent to the next hop right after it is received. Although this guarantees that all the packets are eventually successfully aggregated, it might introduce some performance penalty, because a switch now sends more packets than the optimal.
}

\begin{figure}[h]
    \centering
    \includegraphics[width=\linewidth]{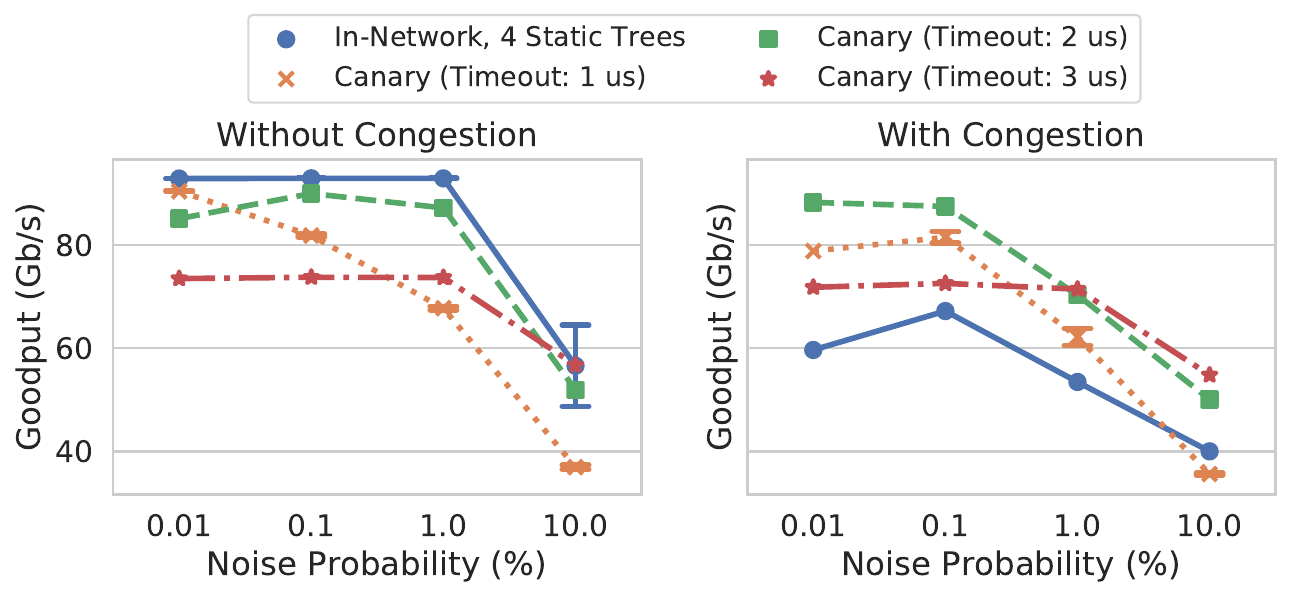}
    \caption{Goodput of a 4MiB allreduce executed on 512 hosts, in a scenario where before sending a packet a host might add a delay of 1us with a given probability.}
    \label{fig:experiments:timeout}
\end{figure}

\revb{
To analyze this scenario, we execute a 4MiB allreduce on 512 hosts, with and without congestion, by comparing it with the in-network allreduce using four static trees and by analyzing the performance for different values of the timeout. Also, every time a host sends a packet, it has a given probability (\textit{noise probability}) of delaying the transmission by 1 microsecond. We report the results of this analysis in Figure~\ref{fig:experiments:timeout}, by showing the runtime when changing the noise probability from 0.01\% to 10\% (i.e., each host on average delays 10\% of the packets it sends by 1 microsecond). 
}

\revb{
When there is no congestion on the network, we observe that \canary is characterized by a lower goodput, as also observed in the previous experiments. Because in this experiment we randomly delay packets by one microsecond, the scenario with a one-microsecond timeout generates several stragglers, decreasing the \canary goodput. This effect is less visible for larger timeouts, which can absorb differences in packet delays.}

\revb{
We also highlight how the performance does not increase nor decrease monotonically with the timeout value. Indeed, a long timeout unnecessarily increases packet latency, whereas a short timeout generates stragglers. However, even if we have a $3x$ difference between the timeout values, we observe at most a $30\%$ difference in the performance. To further mitigate the timeout impact, a possible future extension would be to dynamically select the timeout based on the current network conditions.
}

\revb{
Last, when introducing congestion, \canary instead outperforms the static in-network allreduce regardless of the noise probability and timeout values. Indeed, even if stragglers are generated, their impact on the performance is compensated by the fact packets are forwarded on less-congested paths.
}

\section{Discussion}\label{sec:discussion}
This section discusses some of the limitations of existing programmable switches and their impact on \canary.

\paragraph{Collisions} If two packets with two different \textit{id}s map to the same table entry, \canary forwards the second packet directly to the leader host, generating extra network traffic and potentially reducing the performance (Section~\ref{sec:design:collisions}). For this reason, collisions should happen as rarely as possible. To reduce the number of collisions, in principle \canary could use slightly more sophisticated schemes like Cuckoo hashing~\cite{cuckoo} or double hashing~\cite{GUIBAS1978226}. However, due to the lack of iterative constructs and limited resources, this is not possible on existing programmable switches. As an alternative, the administrator can limit the number of concurrent allreduces and statically partition the table, as done in most in-network allreduce algorithms.

\paragraph{Packet size} Most existing programmable switches can only process a limited number of data elements per packet, based on the number of physical resources available. Although, as discussed in Section~\ref{sec:evaluation:switch}, this number can be increased by exploiting recirculations, it also requires dedicating most of the switch ports to packet recirculations~\cite{switchml,atp}. As an alternative, \canary could be implemented on different programmable switch architectures that do not have limitations on the number of elements that can be processed per packet~\cite{flare}.

\paragraph{Floating-point arithmetic} Most programmable switches do not provide floating-point units~\cite{278324} and, for this reason, most in-network reduction solutions targeting programmable switches assume that the values to be reduced are converted to fixed-point arithmetic before being transmitted over the network~\cite{atp,switchml,omnireduce}. It has been shown that such techniques do not significantly impact the convergence of deep learning training, and thus they could seamlessly be used with \canary. 

\paragraph{Support for other collectives} Although we focused on the allreduce, a similar approach could be used for other collective operations. For example, a \reduce can be easily implemented by selecting as leader node the destination of the \reduce, and by skipping the broadcast phase. Similarly, a \textit{barrier} can be implemented by having a 0-bytes allreduce, and a \textit{broadcast} by having the node acting as the source of the broadcast sending data to the leader host, thus skipping the data aggregation phase. 

\revb{\paragraph{Leader Failure}
Failures of the hosts acting as leaders can be managed with checkpoint/re-start solutions. Indeed, the leader is one of the hosts participating in the allreduce and, if it fails, its data is lost and cannot be recovered as it happens in the case of a switch/link failure. This, however, is also true for any allreduce algorithm (both host-based or in-network) in case of the failure of one of the hosts involved in the reduction. Alternatively, the leader could run on a server not used by any host (e.g., in the SDN controller(s)). However, this would pose scalability challenges in case of multiple in-network allreduce issued by different jobs, and would not allow load balancing between different leaders (see Section~\ref{sec:design:leader}).}

\reva{\paragraph{Fragmentation} We assume that application-level messages are split into IP packets (see Section~\ref{sec:design:general:packets}), each with its own Canary header. We enforce packets (including Canary header) to be no larger than MTU to avoid fragmentation, which would significantly complicate the design.}

\reva{\paragraph{Other Topologies} For the sake of simplicity, we described and evaluated our algorithm on fat tree topologies. However, a similar approach could be used on other topologies, since an aggregation tree is naturally formed when sending packets from the hosts to the root.}

\reva{\paragraph{Retransmission Delays} If we assume a retransmission timeout of $2\cdot RTT$ (where $RTT$ is the \textit{Round Trip Time} between a host and the leader), in the worst case a new reduction for a given block will be re-issued after $3 \cdot RTT$. Indeed, a host needs $2\cdot RTT$ before issuing a retransmission request. The retransmission request arrives at the leader after $RTT/2$, and broadcasts to all the hosts a failure message for that block, that the hosts receive after $RTT/2$.}

\section{Conclusion}\label{sec:conclusion}
In this work, we designed, implemented, and evaluated \canary, the first congestion-aware in-network allreduce algorithm. We first shown the impact that network congestion can have on the performance of in-network allreduce algorithm, up to the point where they exhibit lower performance than host-based allreduce. For this reason, by relying on timeouts, \canary can dynamically route packets to avoid congested links, aggregating them in a best-effort fashion. 

We carefully partitioned \canary functionalities between hosts and switches, and we described a prototype P4 implementation, that we evaluated on a Tofino switch. We then simulated our solution on a 1024 nodes network with 64 switches, showing improvements up to 2x compared to in-network solutions using a single reduction tree and up to 47\% compared to solutions using multiple trees. Furthermore, we have shown that these results are consistent across different congestion intensities, proving that \canary is an effective solution in avoiding congestion when running in-network allreduces. 

\section*{Acknowledgements}
We would like to thank Vladimir Gurevich for the helpful comments and feedback. This work has been partially funded by the European Union’s Horizon Europe programme project RED-SEA (grant no. 955776).
Daniele De Sensi was supported by an ETH Postdoctoral Fellowship (19-2 FEL-50), and by Sapienza University under the SEED-2022 funding scheme.

\bibliographystyle{elsarticle-num}
\bibliography{main}

\end{document}